\documentclass[a4paper,11pt]{article}
\pdfoutput=1 

\usepackage{jcappub} 

\usepackage[T1]{fontenc} 
\usepackage{mathtools}

\def\lsim{\mathrel{\rlap{\lower4pt\hbox{\hskip1pt$\sim$}} \raise1pt\hbox{$<$}}}
\def\gsim{\mathrel{\rlap{\lower4pt\hbox{\hskip1pt$\sim$}} \raise1pt\hbox{$>$}}}
\usepackage{hyperref}
\usepackage{graphicx, color}
\usepackage[export]{adjustbox}

\usepackage[normalem]{ulem}

\newcommand{\newc}{\newcommand}
\newc{\comment}[1]{}

\definecolor{grassgreen}{cmyk}{0.77,0,1,0.05}

\title{Composite Hybrid Inflation : Dilaton and Waterfall Pions}

\author[a]{Giacomo Cacciapaglia,}

\affiliation[a]{IP2I, Universit{\'e} de Lyon, UCBL, UMR 5822 CNRS/IN2P3, 4 rue Enrico Fermi, 69622 Villeurbanne Cedex, France}

\emailAdd{g.cacciapaglia@ip2i.in2p3.fr}

\author[b]{Dhong Yeon Cheong,}

\affiliation[b]{Department of Physics and IPAP, Yonsei University, Seoul 03722, Republic of Korea}

\emailAdd{dhongyeon@yonsei.ac.kr}

\author[a,c]{Aldo Deandrea,}

\affiliation[c]{Department of Physics, University of Johannesburg, PO Box 524, Auckland Park 2006, South Africa}

\emailAdd{deandrea@ip2i.in2p3.fr}

\author[d]{Wanda Isnard,}

\affiliation[d]{ Department of Physics, \'{E}cole Normale Supérieure de Lyon, \\
 46 Allée d'Italie, 69364 Lyon Cedex 07, France}

\emailAdd{wanda.isnard@ens-lyon.fr}

\author[b]{Seong Chan Park}

\emailAdd{sc.park@yonsei.ac.kr}

\abstract{We investigate the possibility that inflation originates from a composite field theory, in terms of an effective chiral Lagrangian involving a dilaton and pions. The walking dynamics of the theory constrain the potential in a specific way, where the anomalous dimensions of operators involving pions play a crucial role. For realistic values of the anomalous dimensions, we find a successful hybrid inflation occurring via the dilaton-inflaton, with the pions acting as waterfall fields. Compositeness consistency strongly constrain the model, predicting a dilaton scale $f_\chi \sim \mathcal{O} (1)$ in unit of the Planck scale, an inflation scale $H_\text{inf} \sim 10^{10}$~GeV, and the pion scale around $10^{14}$~GeV. We further discuss possible phenomenological consequences of this theory.}

\begin{document}
\maketitle
\flushbottom

\section{Introduction}
Cosmic inflation, responsible for the accelerated expansion in our early universe~\cite{Starobinsky:1980te,Guth:1980zm,Linde:1981mu,Starobinsky:1982ee,Albrecht:1982wi,Albrecht:1982mp,Linde:1983gd}, has been pursued for decades (see for example \cite{Linde:2007fr,Baumann:2008bn,kinney2009tasi,Martin:2018ycu} for reviews and detailed bibliography). Pushing the frontier of inflation both theoretically and observationally lead to great progress in understanding the physics involved in this process. However, pinpointing the microscopic origin of the inflation mechanism still remains an open question despite numerous attempts to describe its nature.

In this work, we consider the idea of a composite origin for inflation based on a fundamental gauge theory. Such theories consist of an underlying gauge symmetry that confines at low scales (while being asymptotically free) and that couples to some fundamental fermions. Generally, in the confined phase, such theories contain two classes of ``light'' composite scalars, originating from the spontaneous breaking of global symmetries. The breaking of the chiral symmetry of the fundamental fermions leads to the presence of Nambu-Goldstone bosons (pions). Furthermore, if the theory has a near-conformal dynamics, a light scalar resonance emerges in the form of a dilaton \cite{Yamawaki:1985zg,Matsuzaki_2015}. A constrained potential for the latter is generated by the scale anomaly \cite{Golterman:2016lsd,Appelquist:2017wcg}, and it naturally offers a flat direction useful for generating inflation in the early Universe. Evidence for walking dynamics has been collected both phenomenologically and from lattice data ( see \cite{Cacciapaglia:2020kgq} and references therein). Note that here ``light'' refers to the condensation scale, where other resonances also appear.

Composite inflation has been considered before in the literature, and we refer the reader to these reviews for more complete details \cite{Channuie_2015,Samart}. One early tool that was used to describe a dilaton-like inflaton is holography \cite{Evans_2010}, in the form of a D3/D7 brane system \cite{Dasgupta:2002ew}. In more general composite theories, both mesons \cite{Channuie_2011} and glueballs with dilaton-like dynamics \cite{Bezrukov_2012} have been considered in association to non-minimal coupling to gravity. Another approach involves Nambu--Jona-Lasinio descriptions of the strong dynamics \cite{Channuie:2016iyy,Yuennan_2023}. The possibility of a Nambu-Goldstone inflaton has been explored in \cite{Croon2015GoldstoneI,Croon_2016}. Finally, the dilaton potential from Refs \cite{Golterman:2016lsd,Appelquist:2017wcg} has been considered in Ref.~\cite{Ishida:2019wkd}. In this paper, the inflaton is associated with the dilaton, while pions play a role in determining the effective potential for the inflaton, hence providing an example of two-field inflation.
In this work, we reanalyze and extend the inflaton model of Ref.~\cite{Ishida:2019wkd}, focusing on a consistent origin of the inflaton potential from composite dynamics. In particular, we consider a case where the couplings of the dilaton-inflaton to pions are determined in terms of the scaling dimensions of some composite operators. Hence, realistic values for such scaling dimensions show that new pion couplings are necessary to ensure a timely end of inflation. This finally leads to a waterfall mechanism, triggered by the pions, while inflation occurs along the dilaton direction.
We remark that the couplings of the dilaton to pions may receive other sizable corrections, making them independent from the anomalous dimensions and hence justifying the values chosen, for instance, in Ref.~\cite{Ishida:2019wkd}.


This paper is organized as follows: We first introduce in Sec.~\ref{sec:model} our benchmark theory originating from a composite model, identifying the parameters and their corresponding role. In Sec.~\ref{sec:single} we review the single-field case, extending previous work to understand the impact of the composite scaling dimensions. Finally we consider the complete model in Sec.~\ref{sec:waterfall}, talking into account the waterfall end of inflation due to the pions and imposing consistency conditions to make sure of the composite origin of the potential. This leads to a well-determined and constrained favourable parameter space.
We finally offer our conclusions and discuss the outlook in Sec.~\ref{sec:concl}.


\section{The Model}
\label{sec:model}

For concreteness, we consider a $SU(N_c)$ gauge theory coupled to $N_f$ Dirac fermions in the fundamental representation of the gauge group, even though the effective model can be associated to any underlying gauge dynamics \cite{Cacciapaglia:2020kgq,Cacciapaglia:2022zwt}. For certain values of $N_f$ and $N_c$, the theory will flow to an Infra-Red (IR) fixed point \cite{Dietrich:2006cm,Sannino:2009aw}, hence generating a phase of near-scale-invariant dynamics (walking regime) \cite{Holdom:1981rm,Yamawaki:1985zg}. Further to the IR, fermion and/or gluon condensates can form, leading to the spontaneous breaking of the chiral and scale symmetry. Note that the two symmetries can be broken by the same condensate, or not. The Nambu-Goldstone bosons associated to the breaking of these symmetries are the pions $\phi(x) = \phi^a(x) T^a$ and the dilaton $\chi(x)$, respectively. The pions are usually parameterized in the effective chiral Lagrangian in the non-linear form:
\begin{equation}
    U(x) = \exp \left(  \frac{ i \phi}{f_{\phi}} \right)\,,
\end{equation}
where $f_{\phi}$ denotes the decay constant of the $\phi$ fields. The dynamics of the pions is described by the usual chiral Lagrangian in a derivative expansion, with leading term ${\rm Tr} \left[\partial_{\mu} U^{\dagger} \partial^{\mu} U \right]$. The dilaton couplings, instead, are determined by scale invariance and controlled by a decay constant $f_\chi$, which could be very different from that of the pions $f_\phi$. The dilaton also obtains a potential from the scale anomaly. At low energies, where inflation is expected to take place, the resulting Lagrangian for these composite fields is given by:
\begin{equation} \label{eq:Lagr0}
\begin{split}
    \frac{\mathcal{L}}{\sqrt{-g}} \supset & \frac{1}{2}R- \frac{1}{2} \partial_{\mu} \chi \partial^{\mu} \chi -  \frac{f_{\phi}^2}{2}\left( \frac{\chi}{f_{\chi}} \right)^2 {\rm Tr} \left[\partial_{\mu} U^{\dagger} \partial^{\mu} U \right] - \frac{\lambda_{\chi}}{4} \chi^4 \left( \log \frac{\chi}{f_{\chi}} - A \right) \\
    & + \frac{\lambda_{\chi} \delta_1 f_{\chi}^4 }{2} \left( \frac{\chi}{f_{\chi}} \right)^{3-\gamma_m} {\rm Tr} \left[ U + U^{\dagger} \right]  + \frac{\lambda_{\chi} \delta_2 f_{\chi}^4}{4} \left( \frac{\chi}{f_{\chi}} \right)^{2(3 - \gamma_{4f})} {\rm Tr} \left[( U - U^{\dagger} )^2 \right] \\
    & - V_0\,,
\end{split}
\end{equation}
where we set $M_P =1/\sqrt{8\pi G} = 1$ and the metric at the flat limit $g_{\mu \nu} = (-,+,+,+)$. $V_0$ is a numerical value that sets the vacuum energy to zero.

The first line of Eq.~\eqref{eq:Lagr0} stems from the composite dynamics of the dilaton and pions. In particular, the last term proportional to $\lambda_\chi$ represents the dilaton potential \cite{Golterman:2016lsd,Appelquist:2017wcg}, which is fully determined by the scale anomaly in the theory. The constant $A \sim 1/4$ is chosen so that the dilaton potential has a minimum at $\langle \chi\rangle \equiv \chi_0 = f_\chi$, the dilaton decay constant, which is the order parameter breaking scale invariance. Finally, $\lambda_\chi$ is a dimensionless coupling constant measuring the explicit breaking of scale invariance.

The second line in Eq.~\eqref{eq:Lagr0} contains the effective potential for the pions, generated by explicit breaking terms of the chiral symmetry of the underlying model. For later convenience, we defined the dimensionless coupling constants $\delta_1$ and $\delta_2$ relative to $\lambda_\chi$, even though they are of completely different origin. The first term, controlled by $\delta_1$, is typically generated by a bare mass term for the confining fermions. The second, controlled by $\delta_2$, is generated by four-fermion interactions and, depending on the sign, it could destabilize the minimum of the pion potential away from $\langle \phi \rangle = 0$. The interplay between these two terms has been used in composite Goldstone Higgs models \cite{Cacciapaglia:2014uja}, while here this mechanism can determine the end of inflation via a waterfall potential, as we will see later. In this work, we assume that the couplings of the dilaton $\chi$ to these terms are determined by the scaling dimension of the originating fermionic operators, parameterized in terms of the anomalous dimensions $\gamma_m$ and $\gamma_{4f}$. Their values are fixed by the walking dynamics of the theory, and they can be computed non-perturbatively via lattice results. 
For instance, computations of the mass anomalous dimension $\gamma_m$ in various walking theories seem to suggest that $0 < \gamma_m \lesssim 1$ \cite{DeGrand:2015zxa}. Similar considerations should apply to $\gamma_{4f}$, even though no lattice data is available for this operator. In the following, therefore, we will consider the range
\begin{equation} \label{eq:anomrange}
    0 < \gamma_m\,,\; \gamma_{4f} \lesssim 1,
\end{equation}
as compatible with the composite nature of inflation. However, as it was stressed in Ref. \cite{Ishida:2019wkd}, it is important to note that other scalar mesons in the confined theory or additional contributions from the U$(1)$ axial anomaly could modify the power of the dilaton field $\chi$ in the second line of Eq.~\eqref{eq:Lagr0}. This may destroy the relation to the anomalous dimensions, hence allowing arbitrary exponents in the power-law terms in $\chi$, as considered in Refs~\cite{Iso:2014gka, Kaneta:2017lnj, Ishida:2019wkd}. 

Another consistency relation with compositeness stems from the pion potential in the second line of Eq.~\eqref{eq:Lagr0}. At the minimum for the dilaton potential, $\chi = f_\chi$, the coefficient of these two terms should be small perturbations on the chiral Lagrangian and be determined in terms of the pion decay constant alone, $f_\phi$. Hence, a composite origin of the Lagrangian~\eqref{eq:Lagr0} requires
\begin{equation}
    \lambda_\chi \delta_{1,2} f_\chi^4 \lesssim \mathcal{O} (1)\, f_\phi^4\,.
\end{equation}
The relation above can be expressed as a constraint on the parameter space
\begin{equation}
    {\rm max} (\delta_1,\delta_2) \lesssim \frac{1}{\lambda_{\chi}}\left( \frac{f_{\phi}}{f_{\chi}} \right)^{4}\,,
    \label{eqn:compositebound}
\end{equation}
which we will apply below.
Note that this is based on an order of magnitude estimate, hence it should not be considered as a strict bound but rather as a consistency criterion on the values of the model parameters.


\section{Single Field Inflation and anomalous mass dimension}
\label{sec:single}

We first revisit the single-field case, where the pion minimum is always $\phi=0$ during inflation: this is achieved for $\delta_2 \leq 0$ (we will fix $\delta_2 = 0$ for simplicity). 
Hence, inflation is triggered by the slow-roll of the dilaton $\chi$. This case has been considered in Ref.~\cite{Ishida:2019wkd} but the reference value for the dilaton power was chosen to be $1$, corresponding to $\gamma_m=2$ in the ideal case. As we discussed in the previous section, this is unrealistic from the point of view of composite dynamics, henceforth we will check if a successful inflation can be obtained within the range $0 < \gamma_m \lesssim 1$.

Under the simplifying assumption $\delta_2 = 0$, the inflationary potential becomes
\begin{align}
V(\phi=0,\chi) = - \lambda_\chi \delta_1 f_{\chi}^4 \left( \frac{\chi}{f_{\chi}} \right)^{3-\gamma_m}
+ \frac{\lambda_{\chi}}{4} \chi^4 \left( \log \frac{\chi}{f_{\chi}} - A^{ \rm single} \right) + V_0^{ \rm single}\,,
    \label{potential}
\end{align}
with $A^{\rm single} = \frac{1}{4}  + {\delta_1 (3- \gamma_m)}$ and $V_0^{\rm single} = \frac{1}{16} \lambda_{\chi} f_\chi^4 \left(1 + 4 \delta_1(1+\gamma_m) \right)$. We recall that the values of the two constants stem from the conditions
\begin{equation}
    V_\chi = \frac{\partial V}{\partial \chi} = 0\,, \quad V  = 0\,,
\end{equation}
at the minimum $\langle\phi\rangle=0$ and $\langle\chi\rangle=f_\chi$. The former defines $f_\chi$ as the minimum value of the dilaton field, while the latter sets to zero the vacuum energy.
Note that $\delta_1 = 0$  leads to the pure Coleman-Weinberg (CW) inflation, where the correlation between the e-fold number and the spectral index is given in Ref.~\cite{Barenboim_2014}.

Slow-roll inflation occurs in the small $\chi$-field regime. Using the potential given in Eq.~\eqref{potential}, we compute the usual slow-roll parameters as a function of the field value:
\begin{equation}
	\epsilon_{\chi} = \left.\frac{1}{2} \left( \frac{V_{\chi}}{V} \right)^2\right|_{\phi=0}\,,~~\eta_{\chi} = \left.\frac{V_{\chi \chi}}{{V}}\right|_{\phi=0}\,,~~\xi_{\chi} = \left.\frac{V_{\chi} V_{\chi \chi}}{{V^2}}\right|_{\phi=0}.
\end{equation}
%
%
The inflationary observables can be expressed in terms of these slow-roll parameters:  the spectral index $n_s = 1 - 6 \epsilon_\chi + 2 \eta_\chi$, the tensor-to-scalar ratio $r = 16 \epsilon_\chi$, the spectral index running $\frac{{\rm d} n_s}{{\rm d} \ln k} = -2 \xi_\chi + 16 \epsilon_\chi \eta_\chi - 24 \epsilon_\chi^2$, the curvature power spectrum $\mathcal{P}_{\zeta} (k_*)  = \frac{1}{24 \pi^2} \frac{V_0}{\epsilon_{\chi}}$, and the number of e-folds
\begin{equation}
N_e = \int_{\chi_{*}}^{\chi_\text{end}} {\rm d} \chi \frac{1}{\sqrt{2 \epsilon_\chi}}\,,
\end{equation}
where $\chi_\ast$ is the field value at the pivot scale $k_\ast = 0.05~\text{Mpc}^{-1}$ and $\chi_\text{end}$ at the end of inflation, defined as the field value where the slow-roll parameters approach unity.
The corresponding CMB observations from Planck18+BK18+BAO \cite{Planck:2018jri} give the values
\begin{equation} \label{eq:obs}
\begin{split}
    n_s &= 0.9649 \pm 0.0042, \\
    r &  < 0.036\,, \\
    \ln (10^{10} \mathcal{P}_{\zeta} (k_*)) &= 3.040 \pm 0.0016\,, \\
    \frac{{\rm d} n_s}{{\rm d} \ln k} &= - 0.0045 \pm 0.0067\,.
\end{split} \end{equation}
The CMB pivot scale e-folds is expressed in terms of the potential value during inflation and the reheating temperature as
\begin{align}
    N_e \simeq 61 + \frac{2}{3}\ln \frac{V_0^{1/4}}{10^{16}~\text{GeV}} + \frac{1}{3} \ln \frac{T_{\rm reh}}{10^{16}~\text{GeV}}\,.
\end{align}
The above equation can be used to constrain the number of e-folds in inflationary scenarios.

\begin{figure}[t]
    \centering
\includegraphics[width=0.47\textwidth]{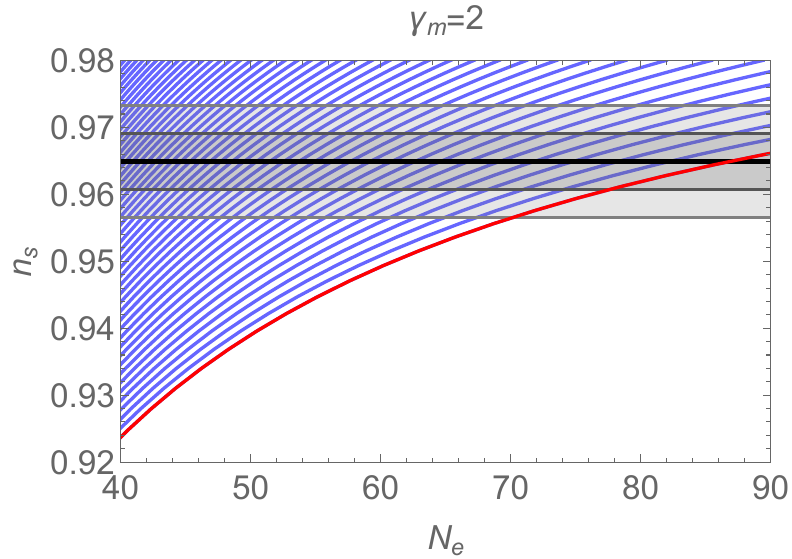}
\includegraphics[width=0.47\textwidth]{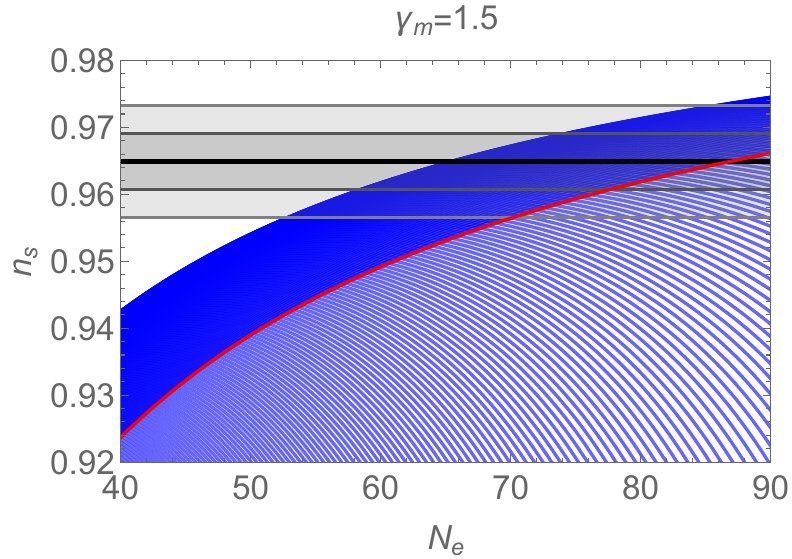}
\includegraphics[width=0.47\textwidth]{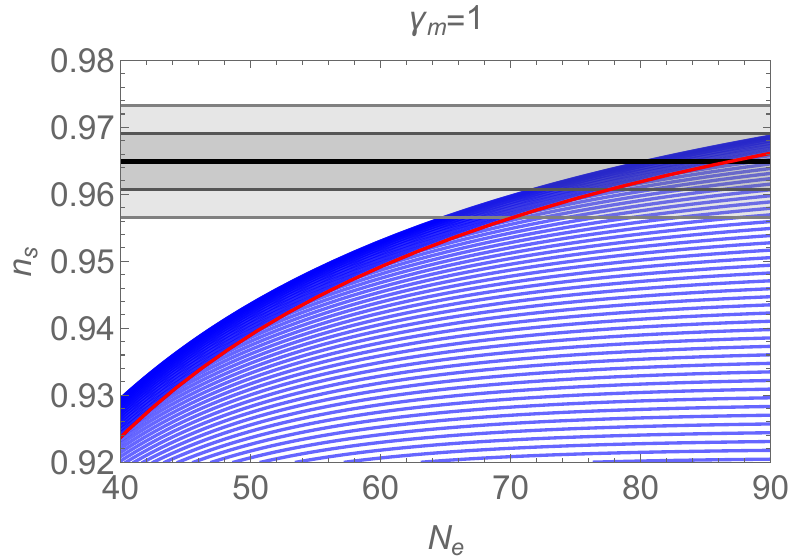}
\includegraphics[width=0.47\textwidth]{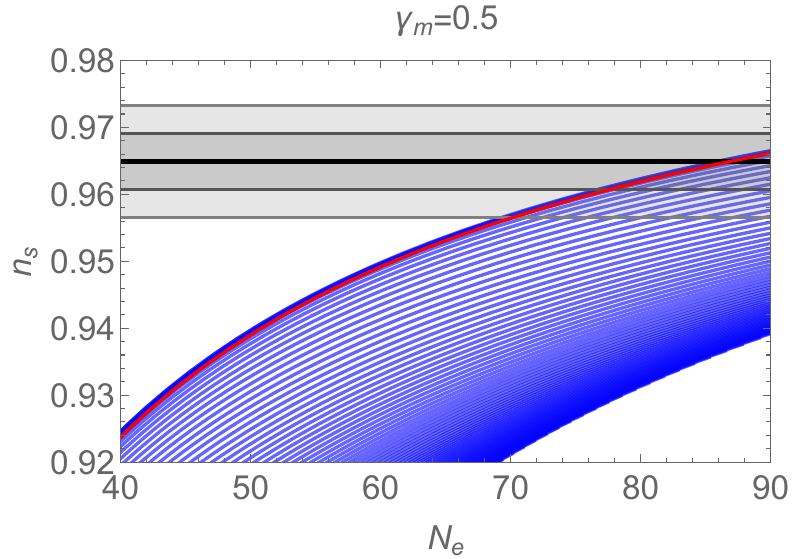}
    \caption{Number of e-folds versus spectral index for a single field slow-roll inflation with the pure Coleman-Weinberg potential (in red, $\delta_1=0$) and varying the coefficient $\delta_1 \in [0,1] \times 10^{-6}$ (in blue) for four different choices of the anomalous mass dimension $\gamma_m=(2,\ 1.5,\ 1,\ 0.5)$. The black line, dark gray and light gray shaded areas correspond to the Planck18+BK18+BAO spectral index central value, $1\sigma$ and $2 \sigma$ contours respectively. $\gamma_m = 2$ allows optimal compatibility with CMB measurements \cite{Ishida:2019wkd}, whereas other values allow marginal $2 \sigma$ compatibility. However, lattice studies suggest $\gamma_m \lesssim 1$, for which inflation is disfavored.}
\label{fig:relation_slow_roll}
\end{figure}

By fixing the anomalous dimension $\gamma_m$ and $f_\chi$, we convert the remaining parameters to the observables $(\mathcal{P}_{\zeta} (k_*), n_s(k_*), N_e(k_*))$. We extract the parameter sets that satisfy  correct CMB power spectrum normalization as in Eq.~(\ref{eq:obs}), then obtain a relation between $n_s(k_*)$ and $ N_e(k_*)$. In Fig.~\ref{fig:relation_slow_roll}, we depict the spectral index $n_s$ as a function of the number of e-folds, with fixed CMB normalization, all evaluated at the CMB pivot scale. The four panels correspond to different values of the mass anomalous dimension $\gamma_m$. We note that for $\gamma_m=2$, good values of the e-fold number can be obtained \cite{Ishida:2019wkd}. As the top-left panel shows, increasing $\delta_1$ reduces the number of e-folding, which is typically too large in the pure CW case. However, for smaller values of $\gamma_m$, this behavior changes and it is not possible to go below a minimal number of e-folds. In particular, the bottom panels show that for values of the anomalous dimension compatible with the compositeness criterion in Eq.~\eqref{eq:anomrange}, i.e. $0<\gamma_m \lesssim 1$, a marginal 2$\sigma$ compatibility is allowed for $\gamma_m \sim 1$. Smaller values of $\gamma_m$ tend to request larger values of e-folds, $N_e \geq 70$, which often leads to unacceptably high reheating temperatures.  

From this analysis we can conclude, therefore, that a single field slow-roll inflation scenario in the case of a composite dynamical origin for the inflaton field is not favored when considering realistic values for the anomalous mass dimension $\gamma_m$.

\section{Hybrid Inflation and Observables}
\label{sec:waterfall}

As the values of the mass anomalous dimension compatible with compositeness are not favored for a single field inflation, we now include the dynamical effect of the pions.
Henceforth, we allow both $\delta_1\neq 0$ and $\delta_2\neq 0$ in this section to study the more general case suggested by composite dynamics.  From Eq.~\eqref{eq:Lagr0}, the potential that we will use for inflation is explicitly given by:
\begin{equation}
\begin{split}
    V(\phi,\chi) = & - \lambda_{\chi} \delta_1 f_{\chi}^4 \left( \frac{\chi}{f_{\chi}} \right)^{3-\gamma_m} \cos \frac{\phi}{f_{\phi}} - \lambda_{\chi} \delta_2 f_{\chi}^4 \left( \frac{\chi}{f_{\chi}} \right)^{2(3-\gamma_{4f})} \sin^2 \frac{\phi}{f_{\phi}} \\
    & + \frac{\lambda_{\chi}}{4} \chi^4 \left( \log \frac{\chi}{f_{\chi}} - A \right) + V_0\,.
    \label{eq:potential}
\end{split}
\end{equation}
From the requirement that the  minimum remains at $\langle\chi\rangle \equiv \chi_0 = f_\chi$ and at zero vacuum energy,
\begin{equation}
    V(\phi_0, \chi_0 ) = 0\,, \text{~~~~~}   V_{\chi}(\phi_0, \chi_0 ) = 0 \text{~~~~~and~~~~~} V_{\phi}(\phi_0, \chi_0 ) = 0 \,,
\end{equation}
one can obtain the expression for $V_0$, $A$ and the vacuum expectation value of the pion fields $\langle\phi\rangle\equiv\phi_0$:
\begin{equation}
    \begin{split}
         V_0
           & = \frac{\lambda_{\chi}f_{\chi}^4}{16} \left[1+2 \frac{ \delta_1^2}{\delta_2} (2  + \gamma_m - \gamma_{4f}) - 8\delta_2 (1-\gamma_{4f})\right]\,, \\
         A & = \frac{1}{4} \left[ 1+  2 \frac{\delta_1^2}{\delta_2} (\gamma_m-\gamma_{4f}) - 8 \delta_2 (3-\gamma_{4f})\right]\,, \\
         \phi_0 & = f_{\phi} \arccos \frac{\delta_1 }{2 \delta_2}\,.
    \end{split} \label{eq:vacuum}
\end{equation}
We request $V_0>0$ to ensure a de-Sitter vacuum. Note also that the non-trivial $\phi_0$ vacuum exists for $0<\delta_1 < 2 \delta_2$~\footnote{For $\delta_1 \geq 2 \delta_2$, the vacuum is stuck at $\phi_0/f_\phi = \pi/2$. Also, for $\delta_2< 0$, the minimum remains at $\phi_0=0$.}.
\begin{figure}[htbp]
    \centering
    \includegraphics[height=5.5cm]{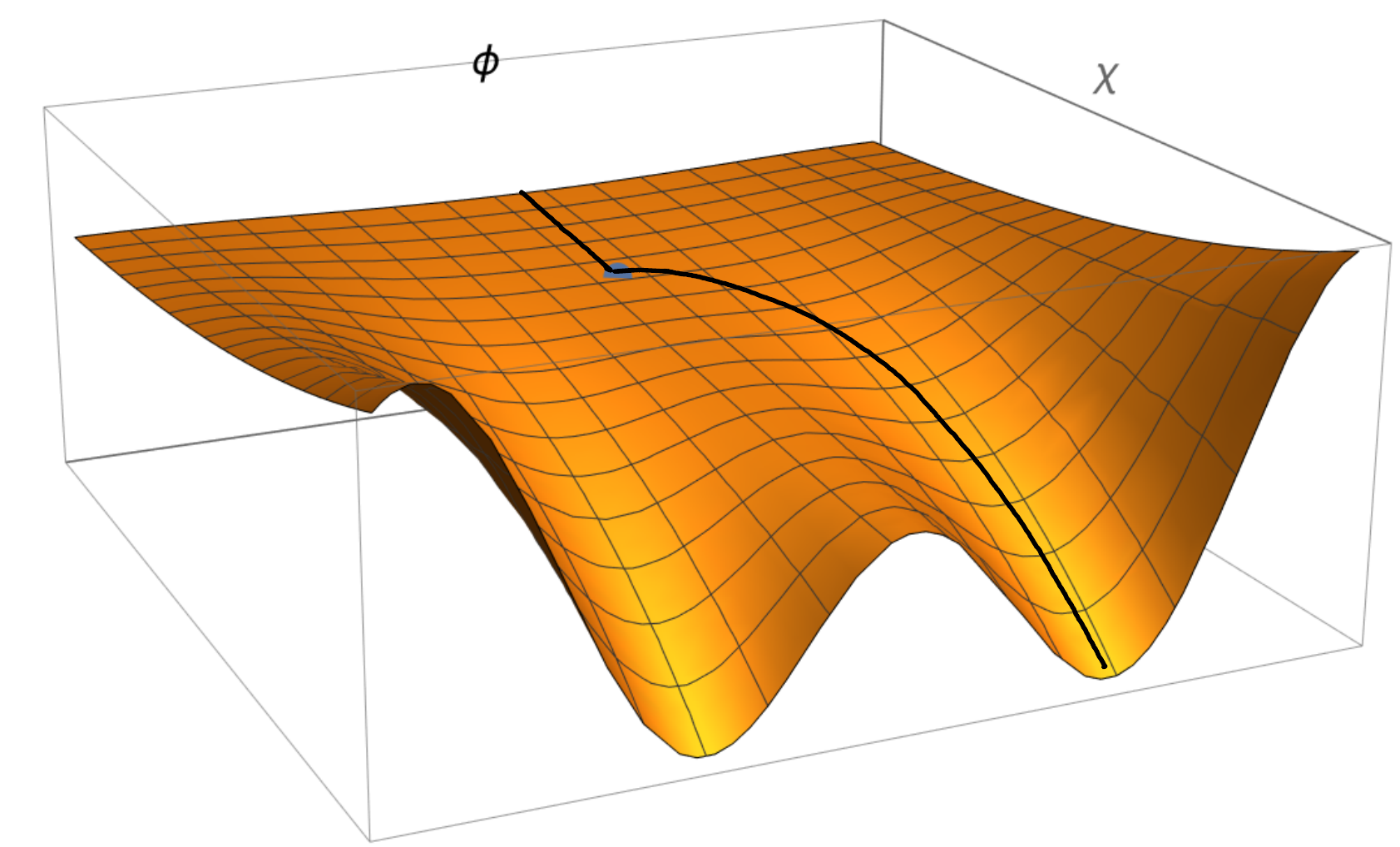}
    \caption{Schematic representation of the inflationary potential, with the black line indicating the trajectory of the fields starting from $\chi_\text{in} \ll f_\chi$ and $\phi_\text{in}=0$. The inflaton field slowly rolls towards $\chi_0 = f_\chi$, however at the turning point (green dot) the pion direction is destabilized.}
    \label{fig:potential}
\end{figure}

Figure~\ref{fig:potential} offers a schematic representation of the inflaton potential. At small $\chi$ values, the $\delta_1$ term dominates the pion potential, hence forcing $\phi=0$. The potential, therefore, consists in a Coleman-Weinberg-like potential in the dilaton $\chi$ direction, where the potential asymptotically goes to a constant value in the small field regime $ V(0, \chi\ll f_\chi) \simeq V_0 $, supporting small-field inflation.
Starting from small dilaton field values, the inflaton rolls down the pion valley, inducing inflation, until a critical value $\chi_c$ where the pions develop  a tachyonic mass. The pions, due to the tachyonic instability, act as waterfall fields terminating inflation and rolling down the waterfall potential, settling into the true vacuum.  
The trajectory along the pion direction is given by:
\begin{equation}
\phi(\chi) = \left \{ \begin{array}{cc}
         0  &\text{when $\chi < \chi_c$}\\
         f_{\phi} \arccos \frac{\delta_1 }{2 \delta_2} \left( \frac{f_{\chi}}{\chi} \right)^{3+\gamma_m-2\gamma_{4f}} &\text{when $\chi > \chi_c$}
    \end{array} \right. ,
\end{equation}
 where a transition ($V_{\phi \phi} $ changes sign) takes place at
 \begin{equation} \label{eq:chiend}
   \chi=\chi_c \equiv \left( \frac{\delta_1}{2\delta_2} \right)^{\frac{1}{3+\gamma_m-2 \gamma_{4f}}} f_{\chi}\,,
   \end{equation}
which is depicted as the sudden turn in the trajectory in Fig~\ref{fig:potential}.  The end of inflation is controlled by $\chi_c \equiv \chi_\text{end}$.
As shown in Fig.~\ref{fig:potential}, the potential in Eq.~\eqref{eq:potential} has two nonequivalent degenerate vacua, hence the formation of domain walls may occur at the end of inflation \cite{DiLuzio:2019wsw}. This can be easily avoided by lifting the degeneracy by additional pion potential terms, like for instance a term linear in $\sin \phi/f_\phi$:
\begin{equation}
    \Delta V = -\lambda_\chi \delta_3 f_\chi^4 \left( \frac{\chi}{f_\chi}\right)^p \sin \frac{\phi}{f_\phi}\,.
\end{equation}
As long as $\delta_3 \ll \delta_{1,2}$, this term will not significantly affect the properties of inflation.

In this model there are 5 parameters that determine the inflationary specifics, namely
\begin{equation}   (\delta_1,~\delta_2,~\lambda_\chi,~f_\chi,~\chi_*),
\end{equation}
with $\chi_*$ denoting the inflaton field value at the pivot scale, $k = 0.05~\text{Mpc}^{-1}$. As $\phi = 0$ during inflation, $\delta_2$ only enters in the determination of the number of e-folds by fixing the end of inflation at $\chi = \chi_\text{end}$. To explore numerically the parameter space of the model, we choose to impose 3 constraints on the following inflationary observables:
\begin{equation}
\mathcal{P}_{\zeta} (k_*) = 2.1\times 10^{-9}\,, \;\;\; n_s = 0.965\,, \;\;\; N_e = 50\,,
\label{eqn:CMB_central_obs}
\end{equation}
chosen to satisfy the experimental observations in Eq.~\eqref{eq:obs} and a consistent duration of inflation.
This allows us to determine the values of $\delta_1$, $\delta_2$ and $\chi^*$, while keeping $\lambda_{\chi}$ and $f_{\chi}$ as free parameters.


\begin{figure}[t!]
\centering
\includegraphics[height=5.4cm]{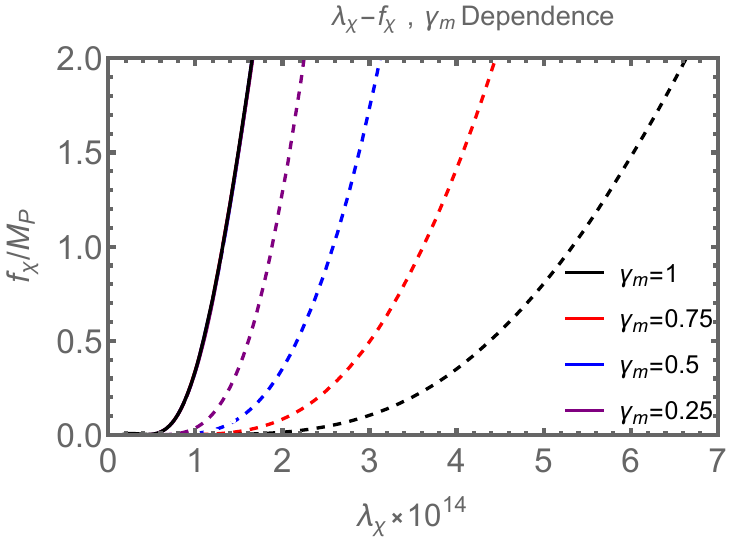} \hfill \includegraphics[height=5.4cm]{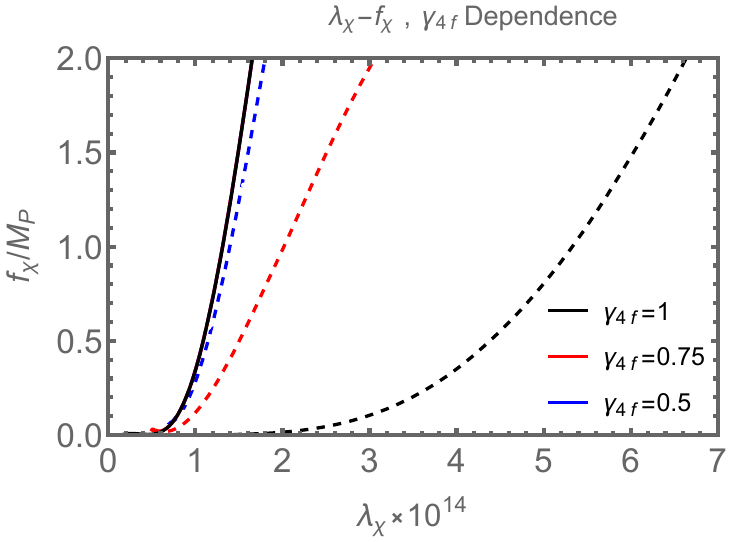}
\caption{ Parameter space dependence on the anomalous mass dimension $\gamma_m$ setting $\gamma_{4f}= 1$ (left) and on the anomalous 4-Fermi dimension $\gamma_{4f}$ setting $\gamma_{m}= 1$ (right). The allowed region corresponds to the regions sandwiched between the solid $(\delta_2 = 10^{-4} )$ and dashed lines $(\delta_2 = 1)$. The region gets restricted by $\delta_2$ values that satisfy $V_0 > 0$ .}
\label{parameterspace3}
\end{figure}

Firstly, we investigate the dependence of the inflation dynamics on the anomalous dimensions $\gamma_m$ and $\gamma_{4f}$. In Fig.~\ref{parameterspace3} we show the allowed values of $f_\chi$ and $\lambda_\chi$, comprised between the solid and dashed lines, for various values of  $\gamma_m$ (left) and $\gamma_{4f}$ (right), while the other anomalous dimension is fixed to 1. By varying the ratio between $\delta_1$ and $\delta_2$, we obtain a parameter set $(\delta_2, \lambda_{\chi}, f_\chi)$ that satisfies Eq.~(\ref{eqn:CMB_central_obs}). We focus on the $\delta_2$ region $0<\delta_2\leq 1$ in order to guarantee that the potential is still dominated by the Coleman-Weinberg like potential, hence the parameter space being confined between a band. We note that for small values of $\delta_2\sim 10^{-4}$, the prediction between $\lambda_\chi$ and $f_\chi$ become independent from the anomalous dimensions. On the other limit, as $\delta_2$ increases, the potential shape differs from the Coleman-Weinberg potential resulting in anomalous dimension dependent predictions.  We remark that reducing the values of the anomalous dimensions from 1 to zero also reduces the allowed parameter space where a successful inflation can occur.





\subsection{Tachyonic Preheating and Waterfall Termination}

In this setup, the mass of the pions takes the following form as a function of the inflaton field $\chi$:
\begin{equation} \label{eq:mphi2}
    m_{\phi}^2 = \frac{\lambda_{\chi} \delta_1 f_\chi^{1+\gamma_m} \chi^{3-\gamma_m} }{f_\phi^2} \cos\left({\frac{\phi}{f_\phi}}\right) - \frac{2 \lambda_{\chi}\delta_2 f_{\chi}^{2(\gamma_{4f}-1)} \chi^{2(3-\gamma_{4f})}}{f_\phi^2}\cos \left({\frac{2\phi}{f_\phi}}\right)\,.
\end{equation}
A hybrid inflation model is realized where the dilaton $\chi$ is the (main) inflaton and the pions $\phi$ are waterfall fields to assist the end of inflation.
In the inflationary valley and at the minimum after inflation, we obtain the simplified relations:
\begin{equation}
    m_{\phi}^2 = \lambda_\chi \frac{f_\chi^4}{f_\phi^2} \times \left \{
    \begin{array}{cc}
         \delta_1 f_\chi^{\gamma_m-3} \chi^{3-\gamma_m} - 2 \delta_2 f_{\chi}^{2(\gamma_{4f}-3)} \chi^{2(3-\gamma_{4f})} &\text{if  $\phi = 0\,,~\chi<\chi_c$}\,, \\
         \left(2\delta_2 - \frac{\delta_1^2}{2\delta_2}\right) &\text{if $\phi= \phi_0\,,~\chi = f_\chi$}\,.
    \end{array} \right.
\end{equation}

During inflation, the waterfall fields are kept at their local minimum $\phi= \phi_0 = 0$, while they become tachyonic when $\chi > \chi_c$. At this point, the perturbations $\delta \phi$ grow exponentially due to the tachyonic instability, leading to a rapid production of pions~\footnote{Tachyonic instability can cause significant density perturbation leading to primordial black hole production and stochastic gravitational wave~\cite{Cheong:2022gfc}. Recently, a hint is suggested from pulsar timing array observations~\cite{NANOGrav:2023hvm}.}. The inflaton rapidly falls down in the $\phi$ direction leading to the end of inflation in the form of tachyonic preheating~\cite{Felder:2000hj,Copeland:2002ku, Gong:2022tfu}. By controlling $\chi_c$ via $\delta_2$, we realize a proper number of e-folds, $N_e\approx 50$.
However, in order to terminate inflation rapidly, the growth of this exponential mode must be compensated by the Hubble expansion. This criterion can be expressed as~\cite{Linde:1993cn}
\begin{align}
    | m_\phi^2 |_{\phi = 0, \chi = \chi_c + \Delta \chi} \gg H^2\,,   \label{eqn:terminationcriteria}
\end{align}
where $\Delta \chi$ corresponds to the field excursion of the dilaton field and $H$ is the Hubble constant at $\chi_c$.

The field excursion is the distance reached in approximately one Hubble time. Recalling that $H^2 \simeq \frac{V}{3} \simeq \frac{V_0}{3}$, the Hubble time is:
\begin{align}
    \Delta t \sim H^{-1} \simeq \frac{4 \sqrt{3}}{  f_{\chi}^2 \sqrt{ \left( \frac{2 \delta_1^2}{\delta_2} (2-\gamma_{4f}+ \gamma_m) + 8 \delta_2 (\gamma_{4f}-1) +1 \right) \lambda_{\chi} }}.
\end{align}
Using the slow-roll approximation $3H\dot{\chi} \simeq - V_\chi $ we can calculate the field excursion:
\begin{align}
    \Delta \chi &\sim -\frac{V_\chi}{3H^2}\simeq - \left. \frac{16 ~\delta_2 ~V_\chi}{\left( 2(2-\gamma_{4f}+ \gamma_m) \delta_1^2 + \delta_2(1+8(\gamma_{4f}-1)\delta_2) \right) \lambda_{\chi} f_\chi^4}\right|_{\phi= 0, \chi = \chi_c }.
\end{align}
As the pion mass squared in Eq.~\eqref{eq:mphi2} is inversely proportional to $f_\phi^2$, the criterion in Eq.~\eqref{eqn:terminationcriteria} will always provide an upper bound on the pion decay constant $f_\phi < f_\phi^\text{waterfall}$. Combined with the compositeness criterion in Eq.~\eqref{eqn:compositebound}, which provides a lower bound $f_\phi > f_\phi^\text{composite}$, we can constrain the values of the pion decay constant compatible with the composite nature of inflation. Note that for some parameters points
\begin{equation}
    f_\phi^\text{composite} > f_\phi^\text{waterfall}\,,
\end{equation}
hence we will deem such points as incompatible with compositeness.


\subsection{Benchmark $\gamma_m = \gamma_{4f} = 1$}

\begin{figure}[t!]
\centering
\includegraphics[height=7cm]{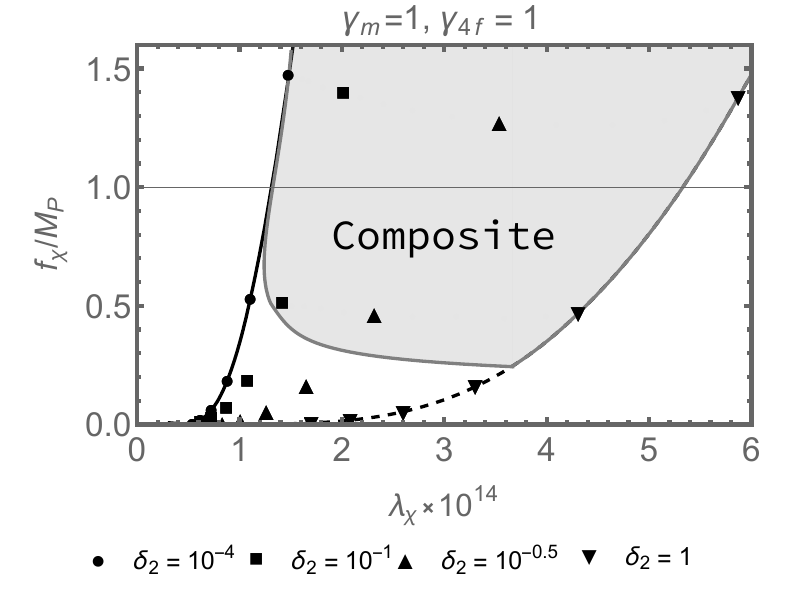}
\caption{\label{fig:cdi_param_space_3} Parameter space allowed considering the CMB observations and $\gamma_m=\gamma_{4f}=1$. The horizontal axis is in units of $\lambda_{\chi} \times 10^{14}$.  The gray region corresponds to the preferred parameter space suggested by the compositeness condition in Eq.~\eqref{compositebound}. }
\label{parameterspace2}
\end{figure}

To be more concrete, we consider here a benchmark with $\gamma_m = \gamma_{4f} = 1$, which are the maximal values of the anomalous dimensions allowed by compositeness. As we showed, this also corresponds to the widest parameter space favorable for inflation. Following the same prescription as in the previous subsection, the favorable parameter space in terms of $\lambda_\chi$ and $f_\chi$ is provided in Fig.~\ref{parameterspace2}.

For $\delta_1 < 2 \delta_2$, as required by the existence of a non-trivial pion vacuum, the mass of the pions reduces to:
\begin{equation}
    | m_\phi^2 |_{\phi = 0, \chi = \chi_c + \Delta \chi} = \frac{4
 \lambda_{\chi} f_{\chi}^2 \delta_1^3 }{f_{\phi}^2 \delta_2^2 } \log \left( \frac{2 \delta_2}{\delta_1} \right)\,.
\end{equation}
The criterion given in the Eq. \eqref{eqn:terminationcriteria} allows us to set the following upper bound on the pion decay constant $f_{\phi}$:
\begin{equation}
    f_{\phi} \ll \frac{8 \sqrt{3} \delta_1^{3/2}}{\delta_2 f_{\chi}} \log^{1/2} \left( \frac{2 \delta_2}{\delta_1} \right)\,.
\end{equation}
Together with the compositeness criterion in Eq.~\eqref{eqn:compositebound}, we obtain a constraint on the two free parameter $f_{\chi}$ and $\lambda_\chi$:
\begin{equation}
    f_{\chi} \lambda_{\chi}^{1/4} \lesssim \frac{8 \sqrt{3} \delta_1^{3/2}}{\delta_2^{5/4} } \log^{1/2} \left( \frac{2 \delta_2}{\delta_1} \right)\,.
    \label{compositebound}
\end{equation}
Only points roughly consistent with this bound are then compatible with a composite origin of the inflationary potential in Eq.~\eqref{eq:potential}. This region is indicated in Fig. \ref{parameterspace2} by the gray area, which mainly requires the dilaton scale $f_\chi$ to remain of the order of the Planck mass.

We further consider a specific benchmark point within the area consistent with compositeness and providing a successful inflation, in order to quantify reasonable values for the pion decay constant~$f_\phi$. As such, we consider the following parameter values:
\begin{equation}
\begin{split}
    & f_{\chi} = 0.5~M_P\,, \;\;\;
    \lambda_{\chi}  = 1.42\times10^{-14}\,, \;\;\;   \chi_*  = 0.004~ M_P\,,   \\
  &  \delta_1 = 10^{-5}\,, \;\;\;
    \delta_2  = 10^{-1}\,;
\end{split}
\end{equation}
for which the pion decay constant must be within the following range
\begin{align}
    9.7\times 10^{-5}~M_P \lesssim f_\phi \lesssim 2.7\times 10^{-4}~M_P\,.
\end{align}
For this particular benchmark point, we can also calculate the Hubble scale at the end of inflation as
\begin{align}
    H_\text{inf} = 4.34\times 10^{-9}~M_P \simeq 10^{10}~\text{ GeV}\,.
\end{align}

Once choosing $f_\phi = 10^{-4}~M_P$, we can also numerically evaluate the masses for the dilaton and the the pions once they relaxed at the true minimum:
\begin{eqnarray}
   \left. m_{\chi}^2 \right|_{\chi = f_\chi, \phi = \phi_0} &=& \lambda_{\chi}f_\chi^2 + 2 \lambda_\chi \frac{ \delta_1^2 f_\chi^2}{\delta_2} \sim 3.57\times 10^{-15}~M_P^2\,, \\
   \left. m_{\phi}^2 \right|_{\chi = f_\chi, \phi = \phi_0} &\sim& 1.8 \times 10^{-8}~M_P^2\,.
\end{eqnarray}
Expressing them in GeV we obtain
\begin{equation}
   m_{\chi} \sim 10^{11}~\text{GeV,~~~~~} m_{\phi} \sim 10^{14}~\text{GeV,~~for~~} f_\phi = 2.44\times 10^{14}~\text{GeV}\,.
\end{equation}

Note that, as the parameter space in Fig~\ref{parameterspace2} is very constrained, the specific numerical value for these parameters cannot vary a lot around the values obtained in this specific benchmark. As we have seen, reducing the anomalous dimensions can further limit the favorable value of the parameters.
In this sense, the composite origin for the inflation model in Eq.~\eqref{eq:potential} \emph{predicts} a composite scale $f_\chi \sim \mathcal{O}(1)~M_P$ and an inflation scale $H_\text{inf} \sim \mathcal{O}(10^{10})~\text{GeV}$, while the pion mass and decay constant are typically a few orders of magnitude below the Planck scale.

\section{Conclusions and Outlook}
\label{sec:concl}


We examined whether a composite theory can trigger cosmological inflation via the dynamical interplay of a dilaton and pions. Inspired by a $SU(N_c)$ model with fermions in the fundamental representation exhibiting a walking dynamics, we impose consistency criteria on the dilaton-pion potential to ensure its composite origin. In particular, the values of the anomalous dimensions, controlling the coupling of the dilaton to pions, are constrained to be smaller than unity as suggested by recent lattice studies. We show that the composite inflation is compatible with the latest Planck18+BK18+BAO observations, only if the pions play the role of waterfall field terminating inflation. The inflaton is played mainly by the dilaton, which slowly rolls from small-field values. The consistency with the composite origin of the potential constrains the parameter space to ensure inflation. 
The inflation scale is around $H_\text{inf} \sim 10^{10}$~GeV, while the pion masses and decay constant are around $f_\phi \sim m_\phi \sim 10^{14}$~GeV. Henceforth, the inflationary dynamics occurs below the Planck scale, in a controllable field-theory regime. The pion potential is also constrained by the inflationary observables, leading to constraints on the interactions that generate such terms by breaking explicitly the chiral symmetry of the composite theory. Hence, the couplings of the Standard Model to the composite sector, once specified, need to respect such constraints.

These stark predictions can lead to interesting consequences for the dynamics of the Universe right after inflation. Indeed, the waterfall end of inflation will inject energy in the pion/dilaton system below the Planck scale. The subsequent reheating of the Standard Model particles will crucially depend on their couplings to the composite sector. For example, some pions may remain stable and contribute to the Dark Matter density, behaving like wimpzillas due to their mass \cite{Kolb:1998ki, Park:2013bza, Rott:2014kfa}. Pion and dilaton decays into the Standard Model may then guarantee the reheating of the visible sector, whose dynamics crucially depends on the allowed strength of their coupling to composite operators, contributing to the inflationary potential. We leave this investigation to future studies. We also note that the turbulent end of inflation could lead to sizable production of primordial black holes and stochastic gravitational waves. Furthermore, the symmetry breaking structure of a specific theory may lead to the production of unstable topological defects, specifically domain walls, which may involve copious production of gravitational waves. Finally, we want to point out that non-minimal couplings between the Ricci scalar and the composite states may lead to possible observational consequences~\cite{Park:2008hz, Kim:2013ehu, Hyun:2023bkf, Hyun:2022uzc}.

\section*{Acknowledgments}

We thank Maria Mylova for helpful discussions. This work was supported by National Research Foundation grants funded by the Korean
government (NRF-2021R1A4A2001897) and (NRF-2019R1A2C1089334) (SCP). GC and AD were partially supported by a Campus France PHC STAR grant. AD acknowledges partial support from the National Research Foundation in South Africa.


\bibliographystyle{JHEP}
\bibliography{ref}

\end{document}